\documentclass[12pt]{iopart}
\usepackage{axodraw,epsfig}
\usepackage{epstopdf}
\usepackage{color}
\usepackage{amssymb}
\usepackage{subfigure}
\def\bea{\begin{eqnarray}}
\def\eea{\end{eqnarray}}
\def\ba{\begin{array}}
\def\ea{\end{array}}
\def\bc{\begin{center}}
\def\ec{\end{center}}
\def\nn{\nonumber}

\def\f{\frac}
\def\a{\alpha}

\def\f#1#2{\frac{#1}{#2}}

\begin{document}

\title[Mixed Bino-Wino-Higgsino Dark Matter in Gauge Messenger Models]{Mixed Bino-Wino-Higgsino Dark Matter \\in Gauge Messenger Models}

\author{Kyu Jung Bae$^1$, Radovan Derm\' \i \v sek$^2$, Hyung Do Kim$^1$ and Ian-Woo Kim$^1$}

\address{$^1$ School of Physics and Astronomy, Seoul National University,
Seoul, Korea, 151-747}
\address{$^2$ School of Natural Sciences, Institute for Advanced Study, Princeton, NJ 08540}
\ead{hdkim@phya.snu.ac.kr, \quad dermisek@ias.edu,\\ 
iwkim@phya.snu.ac.kr, \quad baekj81@phya.snu.ac.kr}
\begin{abstract}
Almost degenerate bino and wino masses 
at the weak scale is one of unique features of gauge messenger models.
The lightest neutralino is a mixture of bino, wino and higgsino
and can produce the correct amount of the dark matter density if  it is the lightest supersymmetric particle.
%(This happens if small contributions to scalar masses from other sources are considered. 
%There is no necessity to rely on special resonance or co-annihilation scenarios. 
%in order to obtain the correct amount of neutralino relic density.
Furthermore, as a result of squeezed spectrum of superpartners which is typical for gauge messenger models, various co-annihilation and resonance regions overlap
and very often the correct amount of the neutralino relic density is generated as an interplay of several processes. 
This feature makes the explanation of the observed amount of the dark matter density much less sensitive to fundamental parameters.
%We consider neutralino of gauge messenger model as a candidate of dark matter in the universe.
We calculate the neutralino relic density  assuming thermal history
and present both spin independent and spin dependent cross sections
for the direct detection.
We also discuss phenomenological constraints from 
$b \to s \gamma$ and muon $g-2$ and  
compare results of gauge messenger models to well known results of the mSUGRA scenario.

\end{abstract}

\maketitle

\section{Introduction}

Models with weak scale supersymmetry are some of the most attractive candidates for extensions of the standard model (SM). Among them the minimal supersymmetric standard model (MSSM) is popular due to its minimality and simplicity. Smallness of the weak scale compared to the Planck scale is nicely explained by the smallness of supersymmetry breaking and the three different gauge couplings meet at the grand unification (GUT) scale, $2 \times 10^{16}$ GeV, which is close to the Planck scale. Furthermore, assuming R-parity, the lightest supersymmetric particle (LSP) is stable and provides a reason for the existence of dark matter.

The lightest neutralino, being weakly interacting, neutral and colorless, and appearing as the LSP in a large class of SUSY breaking scenarios including the most popular one, mSUGRA, is especially good candidate, since it naturally leads, assuming thermal history, to a dark matter density $\Omega_{\rm DM} h^2 \sim 1$~\cite{Jungman:1995df,Bertone:2004pz}.
This observation was certainly a major success of supersymmetry.
However,
the precisely measured value of the dark matter density, $\Omega_{\rm DM} h^2 \sim 0.1 \pm 0.01$~\cite{Spergel:2006hy} together with direct search bounds on superpartners tightly constrain supersymmetric models and explaining the correct amount of the dark matter density while evading all experimental constraints on superpartners is no longer trivial.
For example, a bino-like lightest neutralino which is typical in the mSUGRA scenario usually annihilates too little which results in too much relic density. The bulk region of mSUGRA scenario where neutralino annihilation is 
further enhanced by t-channel exchange of relatively light sleptons and the correct amount of dark matter density can be obtained has been highly sqeezed. Indeed, when the neutralino is mostly bino and $M_1/m_{{\tilde e}_R} \le 0.9$, the correct relic density constrains $m_{{\tilde e}_R} \le 111$ GeV at 95\% CL \cite{Arkani-Hamed:2006mb}.
The limits on the Higgs boson mass and $b \to s \gamma$ independently disfavor the bulk region. In the region with small $\mu$ term neutralinos can efficiently annihilate via their higgsino components. This region extends along the 
line of no EWSB and is refered to as the focus point or hyperbolic branch region. 
Remaining regions are the regions where special relations between independent parameters occure and the neutralino relic density is further reduced by either co-annihilation with other superpartners, e.g. the stau co-annihilation region in mSUGRA when stau mass is very close to the neutralino mass, or by the CP odd Higgs boson resonance when the mass of the CP odd Higgs boson is close to twice the mass of the lightest neutralino. These regions require a {\it critical} choice of parameters in the sense that the predicted value of the dark matter density is highly sensitive to small variations of parameters~\cite{Arkani-Hamed:2006mb}. 

The lightest neutralino in gauge messenger models is typically mostly bino with a sizable mixture of wino and higgsino. The wino and higgsino components enhance the anihilation of the lightest neutralino and the correct amount of the dark matter density is obtained without relying on critical regions of the parameter space.
The virtue of the lightest neutralino being a  mixture of bino and wino was recognized in studies of unconstrained MSSM~\cite{Birkedal-Hansen:2001is,Birkedal-Hansen:2002am,Baer:2005zc,Baer:2005jq}. As already discussed, the bino-like neutralino typically leads to too large relic density. On the other hand both wino-like and higgsino-like LSPs annihilate too fast and 
the correct
 amount of $\Omega_{\rm DM}$ is obtained only if they are very heavy, $m_{\chi^0_1} \sim 1$ TeV for higgsino-like and $m_{\chi^0_1} \sim 2.5$ TeV for wino-like neutralino. Obviously the lightest neutralino which is a proper mixture of bino, wino and higgsino can lead to the correct amount of the dark matter density while avoiding all experimental limits and being fairly light. The only problem is that this situation typically does not happen in widely studied SUSY breaking scenarios. For example, in models with universal gaugino masses at the GUT scale, e.g. mSUGRA, the ratio of bino and wino masses at the weak scale is 1:2 and the sizable bino-wino mixing is not possible. 
However, in gauge messenger models the sizable bino-wino mixing is a built-in feature.  
The bino and wino masses
are generated with the ratio 5:3 at the GUT scale which translates to the ratio 1:1.1
at the EW scale. The bino and wino masses are almost degenerate and thus, besides sizable mixing, also the chargino co-annihilation is always present and playes an important role.

The gauge messenger model is independently well motivated~\cite{Dermisek:2006qj}. The same field which breaks $SU(5)$ to $SU(3) \times SU(2)\times U(1)_Y$, the symmetry of the standard model, is also used to break supersymmetry. 
The heavy X and Y gauge bosons and gauginos play the role of messengers of SUSY breaking.\footnote{This idea was suggested in Ref.~\cite{Dimopoulos:1982gm} motivated by Ref.~\cite{Witten:1981kv}.} All gaugino, squark and slepton masses are given by one parameter and thus the model is very predictive.
Besides already mentioned non-universal gaugino masses at the GUT scale also the squark and slepton masses squared are non-universal and typically negative with squarks being more negative than sleptons. This feature leads to squeezed spectrum at the EW scale (the heaviest superpartner has a mass only about twice as large as the lightest one). Negative stop masses squared at the GUT scale are partially responsible for large mixing in the stop sector at the EW scale which maximizes the Higgs mass and reduces fine tuning in electroweak symmetry breaking~\cite{Dermisek:2006ey}. 
Assuming no additional sources of SUSY breaking the gravitino is the LSP 
 and then, depending on $\tan \beta $, stau or sneutrino is the next-to-lightest SUSY particle (NLSP).
However, 
masses of the lightest neutralino, sneutrino, stau, and stop are very close to each other and thus considering  small additional contributions to scalar masses,
e.g. from gravity mediation the size of which is estimated to be of order 20\% -- 30\% of gauge mediation, neutralino can become the LSP.
In that case we can utilize the bino-wino-higgsino mixed neutralino feature of gauge messenger models to explain the correct amount of the dark matter density. Athough there is no necessity to rely on special resonance or co-annihilation scenarios,  due to the squeezed spectrum of gauge messenger models, these special regions overlap and very often the correct amount of the relic density is generated as an interplay of several processes. 
This feature makes obtaining the correct amount of the dark matter density much less sensitive to fundamental parameters.

In this paper we consider the lightest neutralino of gauge messenger models as a candidate for the dark matter of the universe.
In Sec.~\ref{sec:model} we review basic freatures of gauge messenger models. We discuss neutralino dark matter in mSUGRA scenario in more detail in Sec.~\ref{sec:mSUGRA} which will be useful when comparing results of gauge messenger models. In this section we also outline procedure used to obtain results and summarize experimental constraints used in the analysis. Results of neutralino relic density in gauge messenger models are presented in Sec.~\ref{sec:results} together with the discussion of constraints from $b \to s \gamma$ and muon $g - 2$. We also give predictions for direct dark matter searches.
Finally, we conclude in Sec.~\ref{sec:conclusions}. For convenience, formulae for the composition of the lightest neutralino in gauge messenger models which are used in the  discussion of results
are derived in the Appendix.

\section{Gauge Messenger Model \label{sec:model}}

Let us summarize basic features of  gauge messenger models  introduced in Ref.~\cite{Dermisek:2006qj}.
The simple gauge messenger  model is based on an SU(5) supersymmetric GUT with a minimal particle content. It is assumed that an adjoint chiral superfield, $\hat \Sigma$, gets a vacuum 
expectation value (vev) in both its scalar and auxiliary components: $\langle \hat \Sigma \rangle = (\Sigma + \theta^2 F_\Sigma ) \times \rm{diag} (2,2,2,-3,-3)$. The vev in the scalar component, $\Sigma \simeq M_G$, gives supersymmetric masses to  X and Y gauge bosons and gauginos and thus breaks SU(5) down to the standard model gauge symmetry. The vev in the F component, $F_\Sigma$, splits masses of heavy gauge bosons and gauginos and breaks suppersymmetry. The SUSY breaking is communicated to MSSM scalars and gauginos through loops involving these heavy gauge bosons and gauginos which play the role of messengers (the messenger scale is the GUT scale). The gauge messenger model is very economical, all gaugino and scalar masses are given by one parameter, 
\bea
M_{\rm SUSY} & = & \frac{\alpha_G}{4\pi} \f{|F_\Sigma|}{M_G},
\eea
and it is phenomenologically viable~\cite{Dermisek:2006qj}.

A unique feature of the gauge messenger model is the non-universality of gaugino masses at the GUT scale. The bino, wino and gluino masses
are generated with the ratio 5:3:2 at the GUT scale:
\bea
M_1 & = & 10 \, M_{\rm SUSY}, \label{eq:M1} \\
M_2 & = & 6 \, M_{\rm SUSY} , \\
M_3 & = & 4 \, M_{\rm SUSY} .
\eea
As a consequence, the weak scale bino, wino and gluino mass ratio is approximately 1:1.1:2. 

Similarly, soft scalar masses squared are non-universal and typically negative at the GUT scale. They are driven to positive values at the weak scale making the model phenomenologically viable. 
Negative stop masses squared are a major advantage with respect to the electroweak symmetry breaking which requires less fine tuning and at the same time avoids the limit on the Higgs boson mass by generating large mixing in the stop sector~\cite{Dermisek:2006ey}.
In the gauge messenger model the GUT scale boundary conditions for squark and slepton
masses of all three generations are given as: 
\bea 
m_Q^2 & = & -11 \, M_{\rm SUSY}^2, \label{eq:mQ}
\\
m_{u^c}^2 & = & -4 \,  M_{\rm SUSY}^2,
 \\
m_{d^c}^2 & = & -6 \, M_{\rm SUSY}^2,
 \\
m_L^2 & = & -3  \, M_{\rm SUSY}^2,
 \\
m_{e^c}^2 & = &  + 6 \, M_{\rm SUSY}^2, \label{eq:me}
 \\
m_{H_u,H_d}^2 & = & -3 \, M_{\rm SUSY}^2.
\eea
For completeness, the soft tri-linear couplings are given by:
\bea
A_t & = & -10 \, M_{\rm SUSY}, \\
A_b & = & -8 \, M_{\rm SUSY},  \\            
A_{\tau} & = & -12 \, M_{\rm SUSY} , \label{eq:Atau}
\eea 
and the same results apply to soft tri-linear couplings of the first two generations. The soft SUSY breaking parameters given in Eqs.~(\ref{eq:M1}) -- (\ref{eq:Atau}) correspond to the simple SU(5) gauge messenger model with minimal particle content. For soft SUSY breaking parameters in extended models see Ref.~\cite{Dermisek:2006qj}.

Gauge mediation does not generate the $\mu$ and $B\mu$ terms and they have to be introduced as independent parameters. As we discuss later, they can be generated by gravity mediation through Giudice-Masiero mechanism~\cite{Giudice:1988yz}.
The absolute value of $\mu$ is fixed by requiring EWSB with the correct value of $M_Z$ and thus only the ${\rm sign}(\mu)$ can be chosen arbitrarily. 
The other parameter, $B\mu$, can be replaced by $\tan \beta = v_u/v_d$.
Thus the simple gauge messenger model has two continuous and one discrete parameters:
\bea
M_{\rm SUSY}, \quad \tan \beta, \quad {\rm sign}(\mu).
\eea
Furthermore, constraints on muon anomalous magnetic moment favor the sign of $\mu$
to be the same as the sign of the wino mass which in our notation is positive. 

\begin{figure}
\begin{center}
\subfigure[Simple Gauge Messenger Model]{\includegraphics[width=12.5cm]{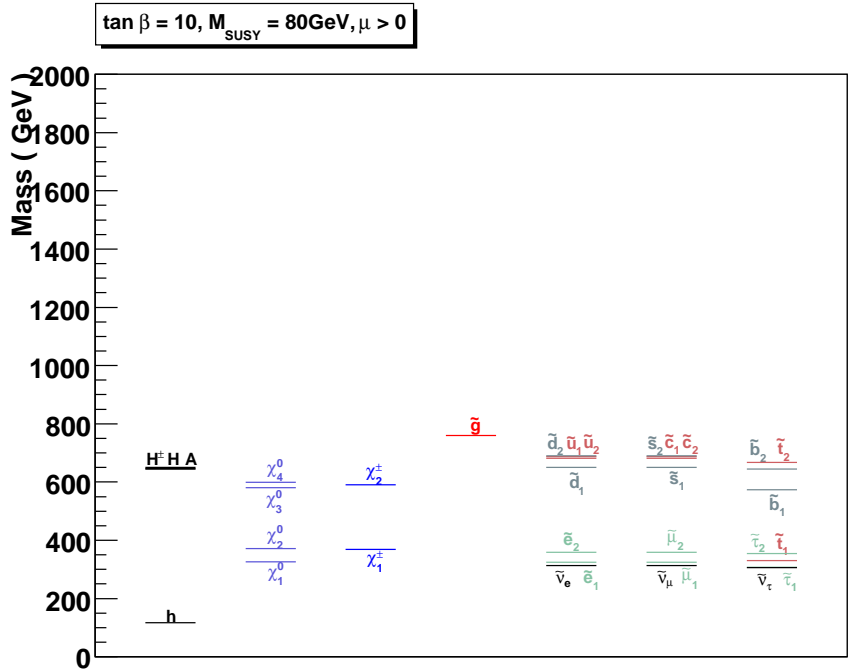}}
%\hspace{0.1in}
\subfigure[mSUGRA]{\includegraphics[width=12.5cm]{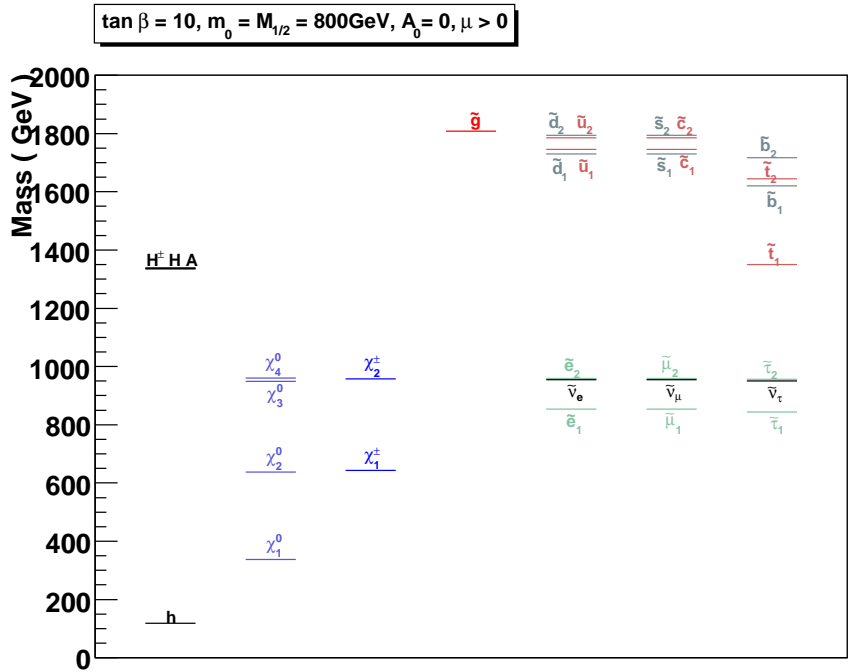}}
\end{center}
\caption{The spectrum of the simple gauge messenger model for $\tan \beta =10$ and $M_{\rm SUSY} = 80$ GeV (a) and the spectrum of mSUGRA for $\tan \beta =10$, $m_0 = M_{1/2} = 800$ GeV, $A=0$ (b).
The parameters are chosen such that the lightest neutralino mass is the same
in both cases.}
\label{fig:spectrum}
\end{figure}

An example of the spectrum of the gauge messenger model is given in Fig.~\ref{fig:spectrum}. For comparison we also give a typical spectrum of the mSUGRA scenario in the same figure.
The mass ratio of the gluino and the lightest neutralino is about 2 in the simple gauge messenger model while it is about 6 in the mSUGRA.
Assuming no additional sources of SUSY breaking the gravitino is the LSP (with the mass of order the EW scale)
 and then, depending on $\tan \beta $, stau or sneutrino is the NLSP.~\footnote{
Neglecting mixing in the stau sector, sneutrino would be the NLSP due to the D-term contribution which is negative for sneutrino and  positive for stau. The mixing in the stau sector is enhanced by $\tan \beta$ and for  $\tan \beta \gtrsim 15$ stau becomes lighter than sneutrino~\cite{Dermisek:2006qj}.}
However, as is we can see in Fig.~\ref{fig:spectrum},
masses of the lightest neutralino, sneutrino, stau, and stop are very close to each other and thus considering  small additional contributions to soft masses,
e.g. from gravity mediation or D-term contributions from breaking of $U(1)$ contained in an extended GUT like SO(10) or E(6), neutralino can become the LSP. In that case we can utilize the bino-wino-higgsino mixed neutralino feature of gauge messenger models to explain the correct amount of dark matter.

Since the messenger scale is the GUT scale, and the gauge
mediation is a one loop effect, the naively estimated size of
gravity mediation induced by non-renormalizable operators
(suppressed by $M_{\rm Pl}$) is comparable to the contribution from 
gauge mediation.
Gauge mediation contribution is given by $M_{\rm SUSY}$,
\bea
 C \f{\a}{4\pi} \left|\f{F}{M_{\rm GUT}}\right|, 
\eea
where C represents group theoretical factors appearing in Eqs.~(\ref{eq:M1}) -- (\ref{eq:Atau}) 
and the contribution from gravity mediation is
\bea
\lambda \left| \f{F}{M_{\rm Pl}} \right|. 
\eea
For a typical $C \sim 5 - 10$ and $\lambda$ of order one we find that the  
gauge messenger contribution is about 5 times larger than the contribution from gravity mediation.

Considering the contribution from gravity, the $\mu$ and $B \mu$ terms can be generated~\cite{Giudice:1988yz}
together with additional contributions to soft masses of the two Higgs doublets which we parameterize by $c_{H_u}$ and $c_{H_d}$
so that soft masses of the two Higgs doublets at the GUT scale are given as:
\bea
m_{H_u}^2 & = & - 3 M_{\rm SUSY}^2 + c_{H_u} M_{\rm SUSY}^2, \\
m_{H_d}^2 & = & - 3 M_{\rm SUSY}^2 + c_{H_d} M_{\rm SUSY}^2,
\eea
In addition we also consider a universal contribution to squark and slepton masses which we parameterize by $c_{0}$ so that, e.g., 
\bea
m_{\tilde{Q}}^2 & = & - 11 M_{\rm SUSY}^2 + c_{0} M_{\rm SUSY}^2,
\eea
and similarly for other squark and slepton masses in Eqs.~(\ref{eq:mQ}) -- (\ref{eq:me}).
Thus in the most general case the parameter space of  gauge messenger models we consider is given by five continuous parameters and the sign of the $\mu$ term:
\bea
M_{\rm SUSY}, \quad \tan \beta, \quad c_0, \quad c_{H_u} ,  \quad c_{H_d}, \quad {\rm sign}(\mu).
\eea
Small contribution from gravity mediation, $c_{0} > 5$,  is enough to make neutralino lighter than sneutrino or stau 
in most of the parameter space. Neutralino is then the LSP or NLSP depending on the gravitino mass. Making gravitino heavier is not problematic and it can be done assuming other sources of SUSY breaking which do not contribute to soft SUSY breaking terms of the MSSM sector.
In the next section we consider neutralino LSP as a candidate for dark matter.

\section{Neutralino dark matter in mSUGRA \label{sec:mSUGRA}}

\begin{figure}[t]
  \begin{center}
  %\subfigure[mSUGRA with $\tan \beta=10$]
  \includegraphics[width=12.5cm]{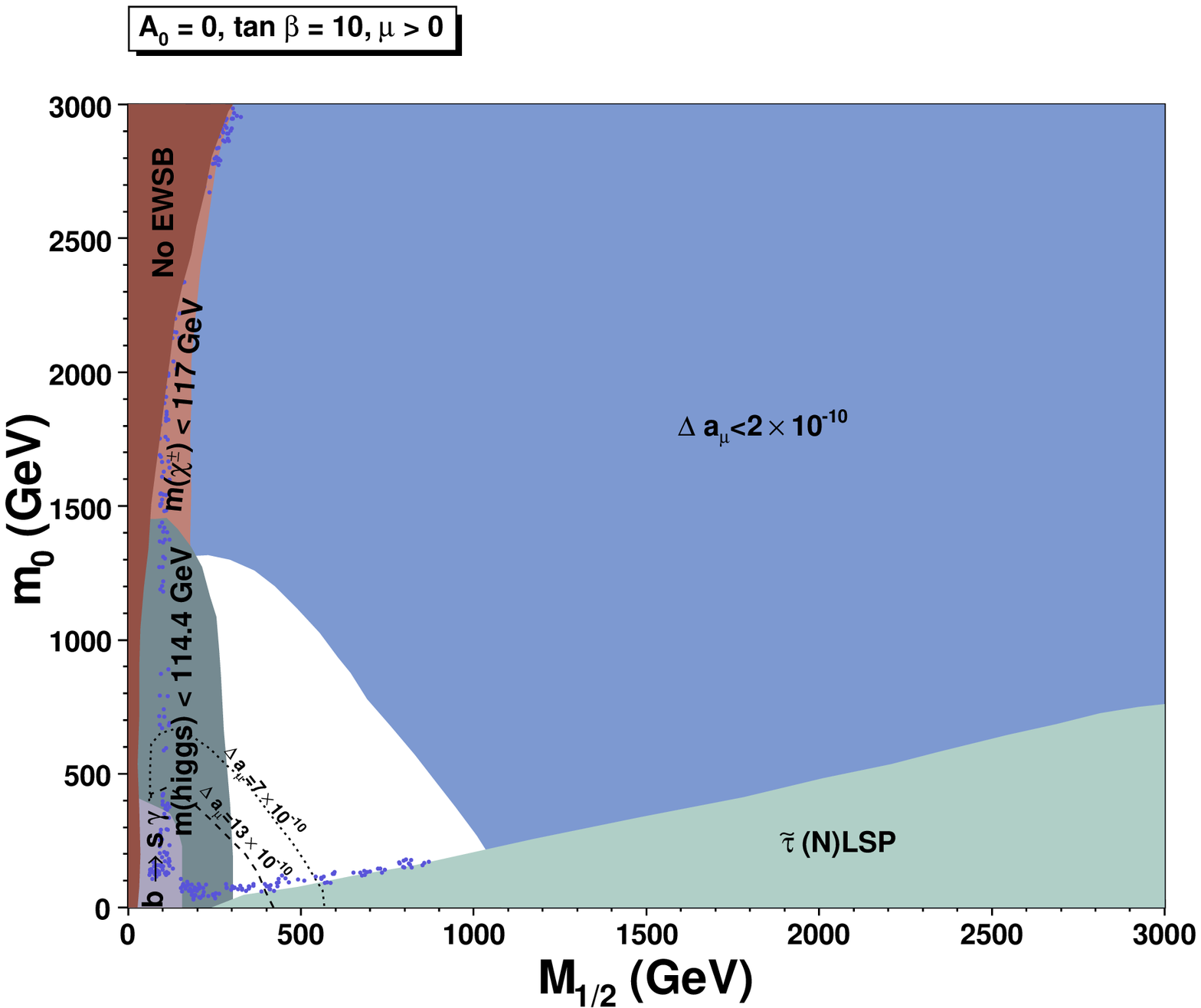}
  %\subfigure[mSUGRA with $\tan \beta =50$]
  %{\label{fig:mSUGRAtan50}
  \includegraphics[width=12.5cm]{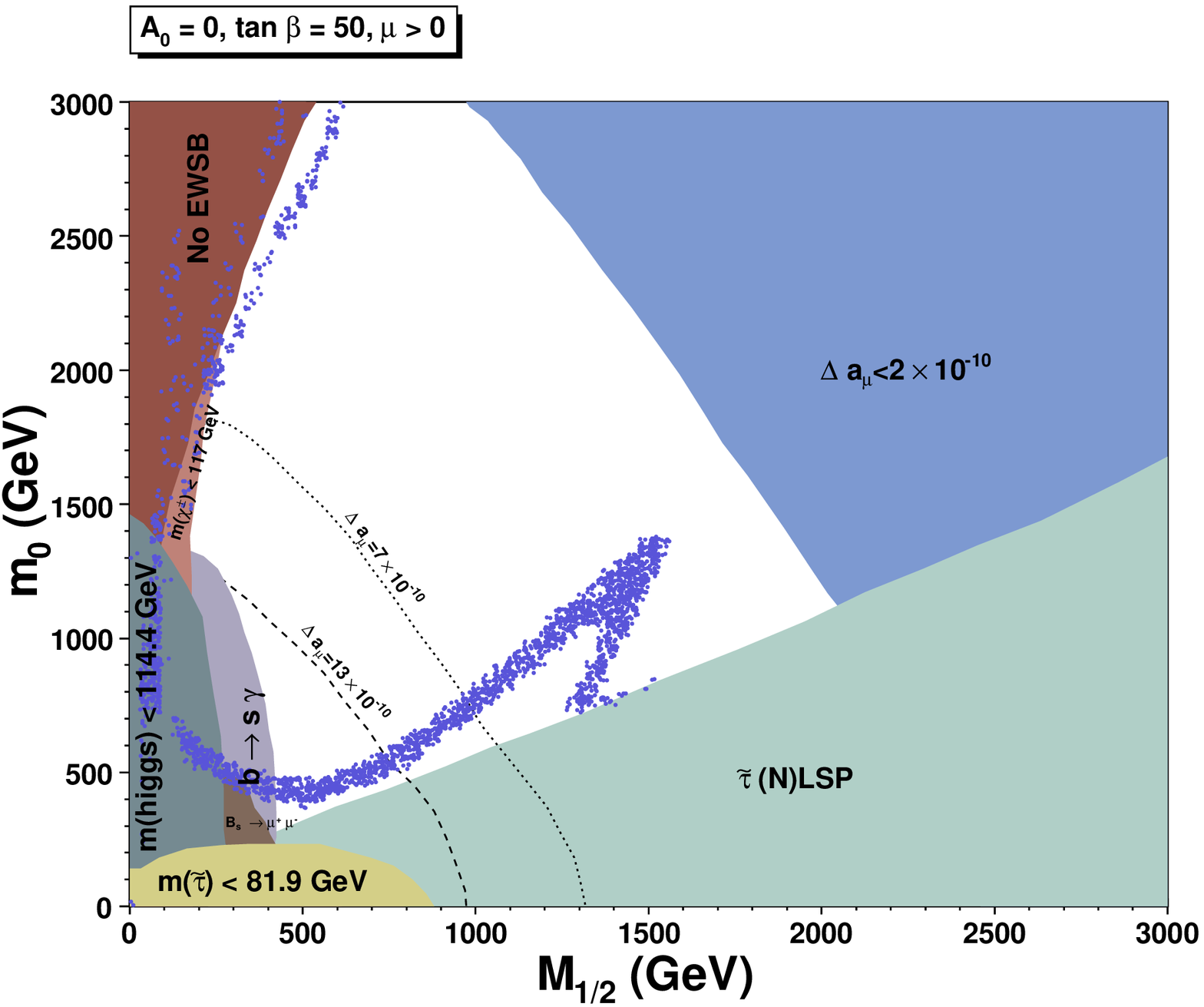}
  \end{center}
\caption{\label{fig:mSUGRA}
Slices through the parameter space of  mSUGRA in $m_0$ -- $M_{1/2}$ plane for 
$\tan \beta =10$ (up) and $\tan \beta=50$ (down) with $A_0 = 0$ and $\mu > 0$.
Blue dots represent the region in which the neutralino relic density is within WMAP range. Shaded regions are excluded by various constraints.}
\end{figure}

In the mSUGRA scenario, or in general in any model with universal gaugino masses at the GUT scale, the lightest neutralino is a mixture of  bino and  higgsino. The  bino-like neutralino typically has a very small annihilation cross section and can not annihilate efficiently. As a consequence, if neutralino is the LSP it gives too large relic density and thus most of mSUGRA parameter space is ruled out by WMAP data. Representative slices through mSUGRA parameter space are shown in Fig.~\ref{fig:mSUGRA} for $\tan \beta = 10$ and $\tan \beta = 50$. The white region represents allowed region after various constraints are imposed (these are indicated in the plots and will be discussed later) and the blue dots represent the region in which the neutralino relic density is calculated to be within the WMAP range.
In several specific regions of parameter space the neutralino relic density is further reduced and these regions are compatible with WMAP data.  In the ``bulk region" (small $m_0$ and $M_{1/2}$)  neutralino annihilation is enhanced by t-channel exchange of sleptons. This region is however disfavored by the limit on the Higgs boson mass and $b \to s \gamma$.  
Contrary to the bino-like neutralino, the Higgsino-like neutralino annihilates too efficiently and the relic density turns out to be smaller than the WMAP value.
When the bino mass and the $\mu$ term are comparable, sizable mixing is possible and the correct relic density is obtained.
This is the region in Fig.~\ref{fig:mSUGRA} for larger $m_0$ which goes along the line where EWSB is no longer possible (or chargino mass limit). In the region where the stau mass is close to the neutralino mass the neutralino relic density is reduced by co-annihilation with stau.
For $\tan \beta =10$, only tiny stau co-annihilation region is available. Bulk region and the focus point region is excluded by direct search bound on the lightest Higgs and chargino and the muon anomalous magnetic moment.  It is indeed the case that the most of the parameter space producing the correct dark matter density is already ruled out. For $\tan \beta = 50$, in addition to the stau co-annihilation region, the funnel region (pseudoscalar Higgs resonance) appears and  also large part of mixed Higgsino region (focus point) is not ruled out for larger $m_0$.

\begin{figure}[t]
  \begin{center}
  \includegraphics[width=12.5cm]{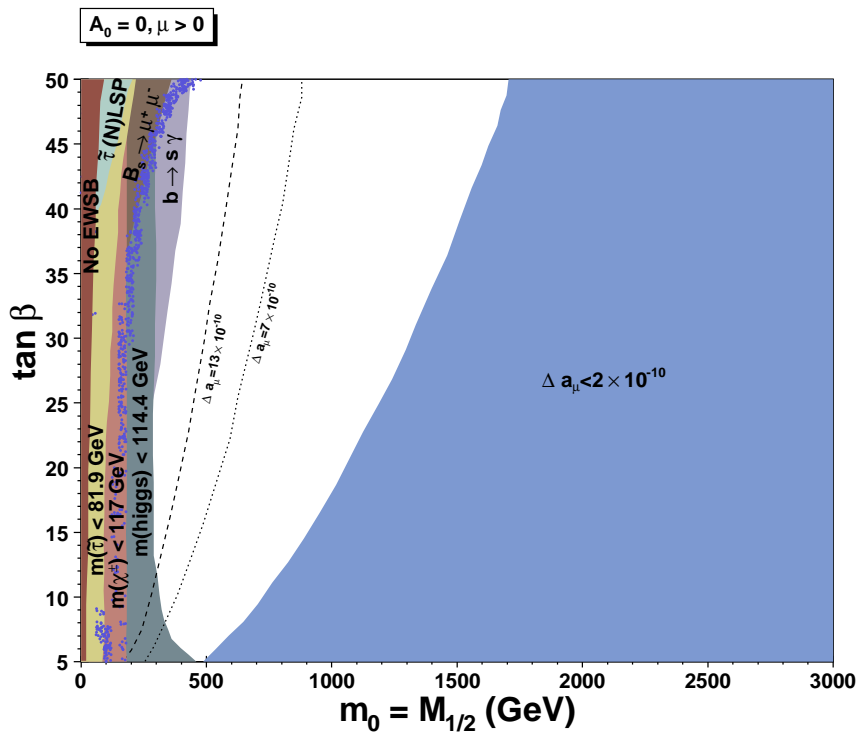}
  \end{center}
\vspace{-0.5cm}
\caption{ \label{fig:mSUGRA_m0=m12}
A slice through the parameter space of mSUGRA in $\tan \beta$ -- $m_0 = M_{1/2}$ plane for 
$A_0 = 0$ and $\mu > 0$.
Blue dots represent the region in which the neutralino relic density is within WMAP range. Shaded regions are excluded by various constraints.}
\end{figure}

In order to compare the result with gauge messenger models that we will discuss later, we fix the ratio of $m_0/M_{1/2}$ and present a slice through mSUGRA parameter space in $m_0$ (or $M_{1/2}$) -- $\tan \beta$ plane in Fig.~\ref{fig:mSUGRA_m0=m12}. The region producing the correct relic density exists only for large $\tan \beta$ ($\tan \beta \ge 45$) which is due to the pseudoscalar Higgs resonance.

\subsection{Experimental constraints and procedure used \label{sec:constraints}}

Before we discuss results for gauge messenger models let us summarize the procedure we use to calculate the neutralino relic density and experimental constraints we employ.
We obtained our results using SOFTSUSY 2.0.8~\cite{Allanach:2001kg} for the renormalization group evolution of soft SUSY breaking parameters from the GUT scale to the EW scale and for calculation of SUSY spectrum. 
The mass of the lightest CP even Higgs boson is calculated with FeynHiggs 2.4.1~\cite{Heinemeyer:1998yj}.
Indirect constraints from $b \rightarrow s \gamma$, muon $g-2$ and $B \rightarrow \mu^+ \mu^-$, and the neutralino relic density are obtained using
micrOMEGAs 2.0~\cite{Belanger:2006is} and the direct detection rates are 
calculated using DarkSUSY 4.1~\cite{Gondolo:2004sc}.

The WMAP result for the dark matter density is~\cite{Spergel:2006hy}:
\bea
\Omega_{\rm DM} h^2 \sim 0.113 \pm 0.009.
\eea
In our plots we consider $2 \sigma$ bounds for the neutralino relic density,  $0.09 \le \Omega_{\rm DM} h^2  \le 0.13$, to be in agreement with the observed dark matter density. This region is represented by blue bands in plots.

For $ B(b \rightarrow s \gamma)$ we consider the allowed range to be
$2.3 \times 10^{-4} \le B(b \rightarrow s \gamma) \le 4.7 \times 10^{-4}$ which is obtained by summing the experimental and theoretical error linearly and taking the $2 \sigma$ range~\cite{Barberio:2006bi,Gambino:2001ew}. 

Muon anomalous magnetic moment might be the only indirect evidence for the presence of new physics at around the weak scale. Recent experimental value of $a_{\mu} = (g-2)_{\mu} /2$ from the Brookhaven "Muon g-2 Experiment" E821 \cite{Bennett:2004pv} is
\bea
a_{\mu}^{\rm exp} & = & ( 11\ 659\ 208 \pm 5.8 ) \times 10^{-10}.
\eea
The standard model prediction contains QED, EW and hadronic parts. As a result of
undertainties in hadronic contribution, we quote results of two groups for $\Delta a_{\mu} = a_{\mu}^{\rm exp} - a_{\mu}^{\rm th}$ where $a_{\mu}^{\rm th}$ stands for the theoretical prediction of the standard model.
From results of Refs.~\cite{Hagiwara:2003da} and \cite{Knecht:2001qg} we have
\bea
\Delta a_{\mu} & = & (31.7 \pm 9.5) \times 10^{-10},
\eea
which indicates a $3.3 \sigma$ deviation from the standard model.
In order to explain the experimental result within $2 \sigma$, we need a contribution from new physics $\Delta a_{\mu} \ge 13 \times 10^{-10}$.
On the other hand, from results of  Refs.~\cite{Davier:2003pw} and \cite{Melnikov:2003xd}, we have
\bea
\Delta a_{\mu} & = (20.2 \pm 9.0) \times 10^{-10},
\eea
which indicates a $2.1 \sigma$ deviation. In this case we need $\Delta a_{\mu} \ge 2 \times 10^{-10}$ if we allow for $2 \sigma$ variation.
Both groups calculated the hadronic contribution using $e^+ e^-$ data. The $\tau$ decay data has not been used because of the uncertainties related to isospin breaking effects.
By combining these two results \cite{Hocker:2004xc,Heinemeyer:2004gx}, we get
\bea
\Delta a_{\mu} & = & (25.2 \pm 9.2) \times 10^{-10},
\eea
which indicates a $2.7 \sigma$ deviation from the standard model. A contribution from new physics 
 $\Delta a_{\mu} \ge 7 \times 10^{-10}$ is necessary in this case to agree with data.

In our plots, we draw all three $2 \sigma$ bounds, $\Delta a_{\mu} = (2,7,13)\times 10^{-10}$. As we neglected $\tau$ decay data, we take the most conservative bound, $\Delta a_{\mu} = 2 \times 10^{-10}$,  to constrain the parameter space.
For illustrative purposes we add two dashed lines corresponding to $\Delta a_{\mu} = 7 \times 10^{-10}$ and $\Delta a_{\mu} = 13 \times 10^{-10}$.

\section{Neutralino dark matter in gauge messenger models \label{sec:results}}

The squeezed spectrum of gauge messenger models makes the discussion of the neutralino relic density very complex.
Various regions with correct relic density which are usually well separated in scenarios with highly hierarchical 
spectrum are overlaping here and often there is no single process that
would be crucial for obtaining the correct amount of the neutralino relic density.

The lightest neutralino in gauge messenger models is typically mostly bino with a sizable mixture of wino and higgsino. In order to understand the dependence of the neutralino relic density on fundamental parameters it is important to know the composition of the lightest neutralino.
The formulae for wino and higgsino components of the lightest neutralino mass eigenstate are derived in the Appendix
and for $\tan \beta \ge 10$  they can be writen as:
\bea
N_{11} & \simeq & 1, \\
N_{12} & \simeq & -\f{M_Z^2 \sin 2\theta_W}{2\epsilon (\mu^2 - M_1^2)}, \\
N_{13} & \simeq & \f{\mu M_Z \sin \theta_W }{\mu^2 - M_1^2}, \label{eq:N13}\\
N_{14} & \simeq & -\f{M_1 M_Z \sin \theta_W }{\mu^2 - M_1^2},\label{eq:N14}
\eea
where $\epsilon$ is defined as $M_2  =  M_1 (1 + \epsilon)$. The bino/wino mass ratio is fixed in the gauge messenger model. As $M_i/g_i^2$ is RG invariant at the 1-loop level, this  ratio at the EW scale is  
\bea
\frac{M_1(M_Z)}{M_2(M_Z)} & = &  \frac{5}{3} \tan^2 \theta_W \frac{M_1(M_{\rm GUT})}{M_2(M_{\rm GUT})}  
\simeq   0.9,
\eea 
which means $\epsilon \simeq 0.1$.
From the above equations we see that the wino and higgsino mixing is sizable unless the ratio $M_Z/\mu$ is too small. For $\mu \ge M_1$, the down type Higgs component, $N_{14}$,
is larger than the up type Higgs component, $N_{13}$. The bino-wino mixing, $N_{12}$ 
is suppressed compared to the bino-higgsino mixing by $M_Z/\mu \le 1$. This is the reason why the bino-wino mixing is negligible in most of SUSY breaking scenarios. However, in gauge messenger models the mixing is enhanced by $1/\epsilon$ thanks to near degeneracy of bino and wino.
As a result, the lightest neutralino in gauge messenger models is mostly bino with sizable and comparable wino and higgsino components.

\begin{figure}[t]
  \begin{center}
   \hspace{-.5cm}
  \includegraphics[width=12.5cm]{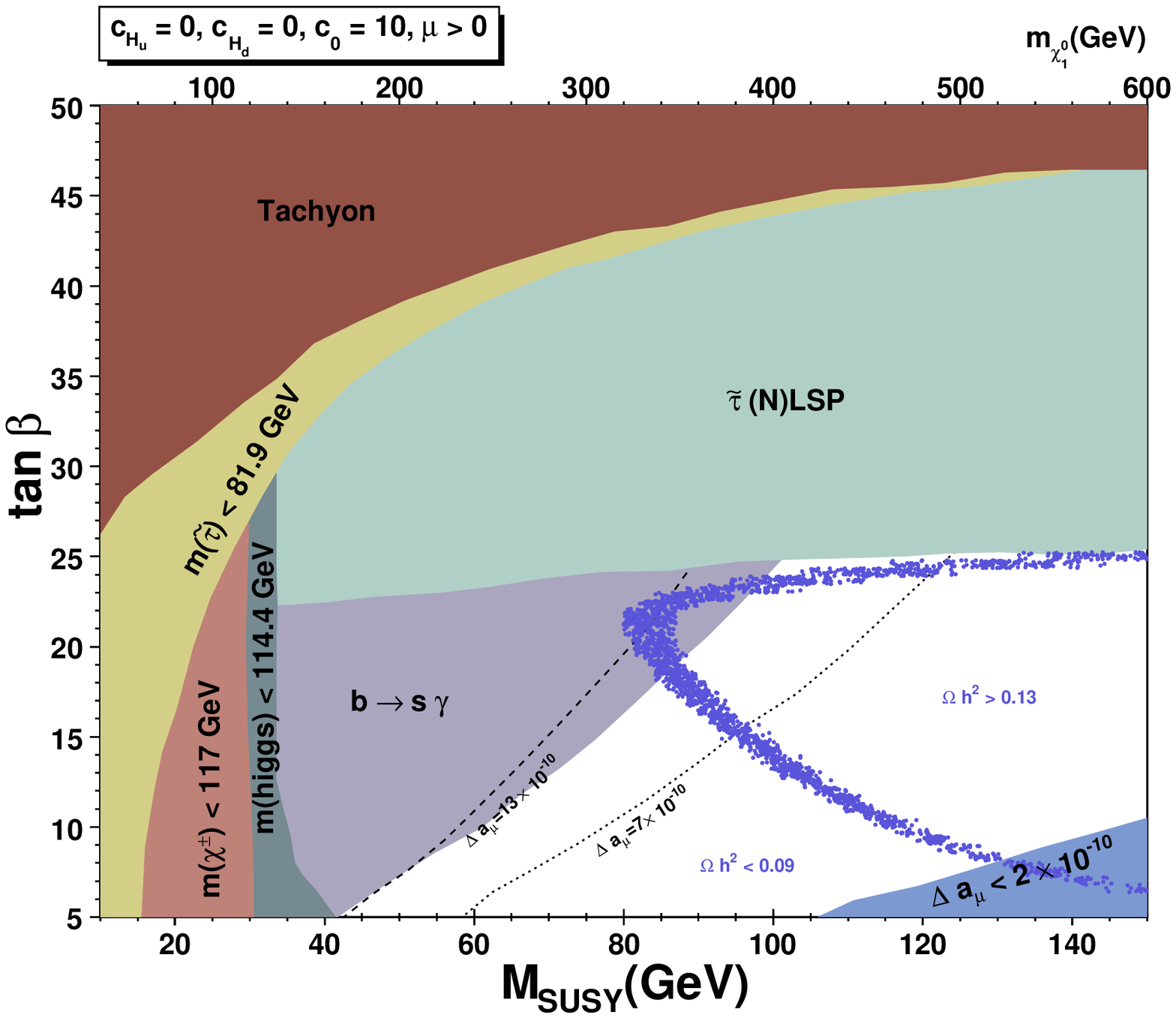}
\hspace{-0.5cm}
  \includegraphics[width=12.5cm]{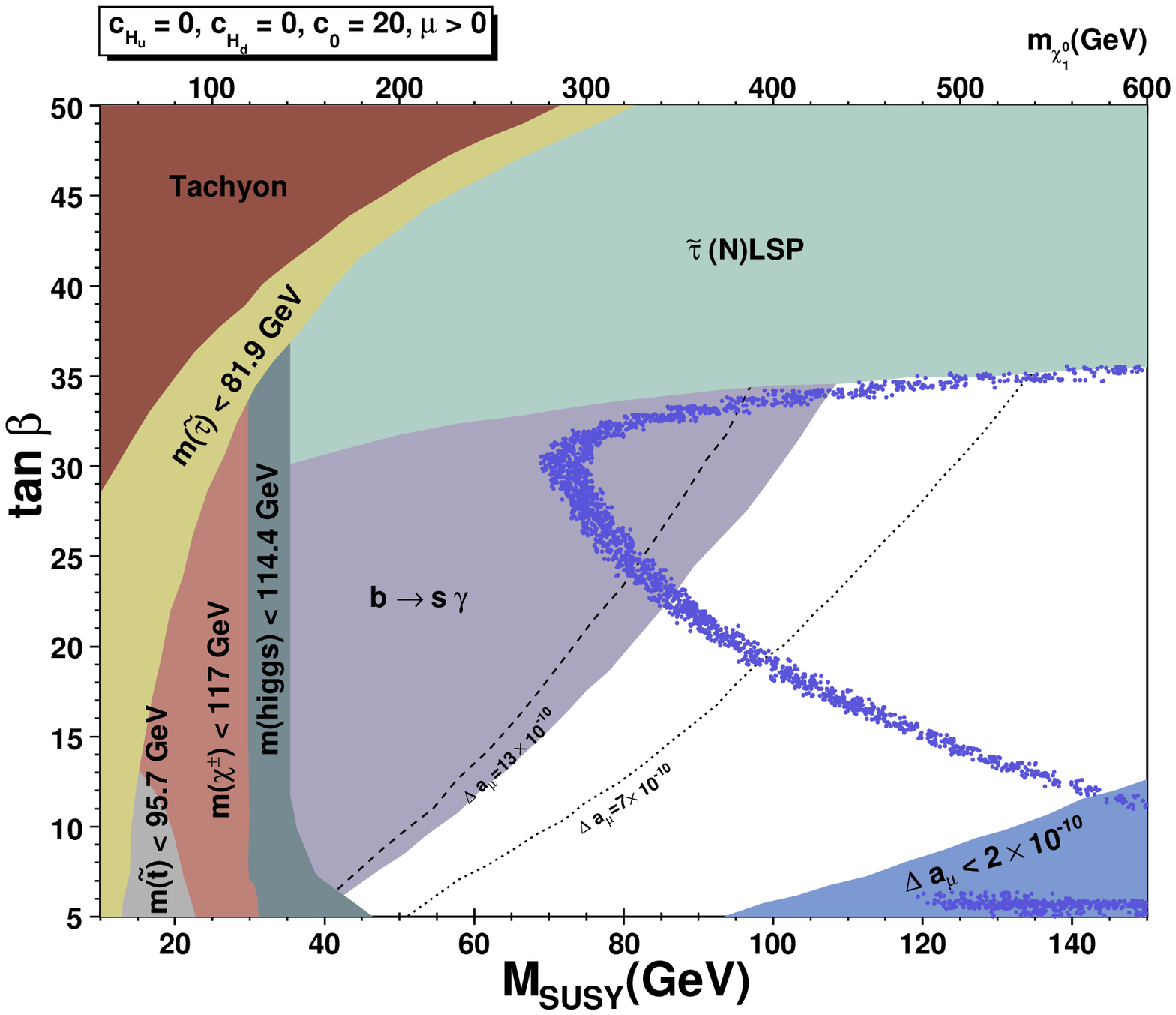}
  \end{center}
\vspace{-0.5cm}
\caption{ \label{fig:0a_0a'} 
The neutralino relic density in $M_{\rm SUSY}$ -- $\tan \beta$ plane of the gauge messenger model 
with $c_{H_u}=c_{H_d}=0$, $\mu > 0$ and
$c_0 = 10$ (up) and $c_0 = 20$ (down).
Blue dots represent the region in which the neutralino relic density is within WMAP range. Regions with too much or too little relic density are indicated. For convenience, the top axis indicates the mass of the lightest neutralino.
Shaded regions are excluded by various constraints. }
\end{figure}

\begin{figure}[t]
  \begin{center}
  \includegraphics[width=12.5cm]{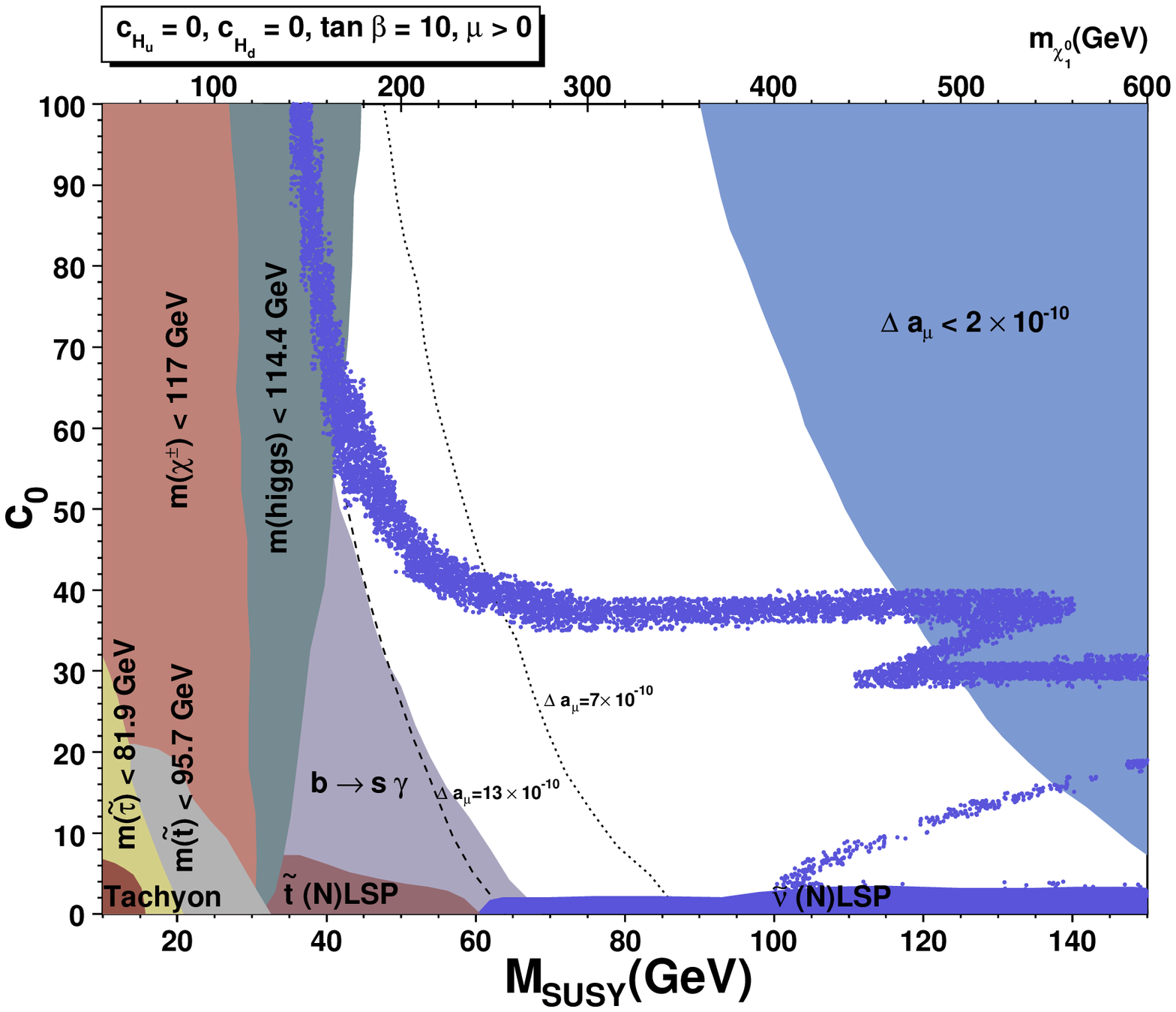}
  \end{center}
\vspace{-0.5cm}
\caption{ \label{fig:0b} 
The neutralino relic density in $M_{\rm SUSY}$ -- $c_0$ plane of the gauge messenger model 
with $c_{H_u}=c_{H_d}=0$, $\mu > 0$ and $\tan \beta = 10$. .
Blue dots represent the region in which the neutralino relic density is within WMAP range. 
Shaded regions are excluded by various constraints. }
\end{figure}

\begin{figure}[t]
  \begin{center}
     \hspace{-.5cm}
  \includegraphics[width=12.5cm]{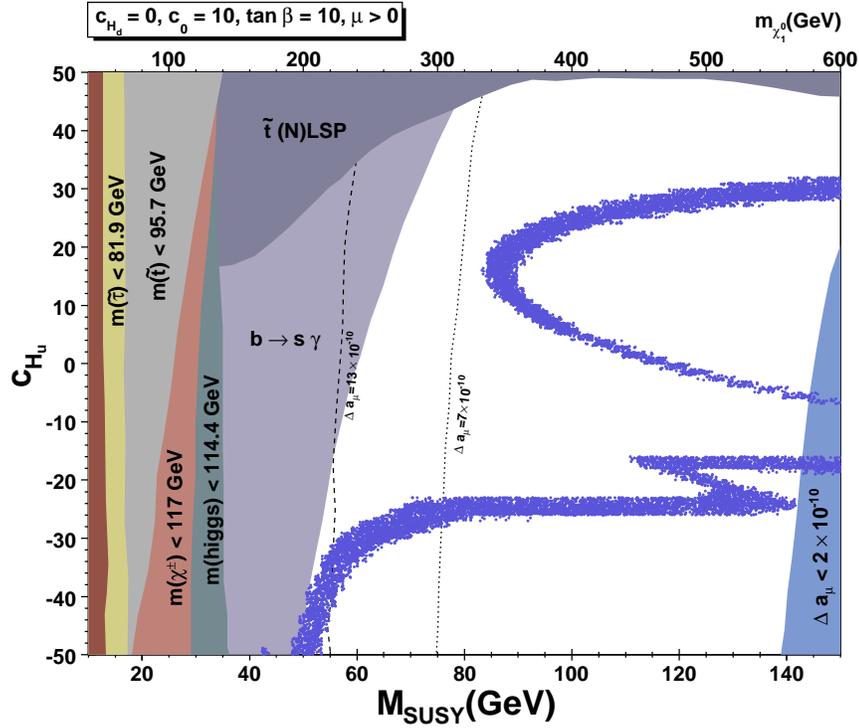}
 \hspace{-0.5cm}
  \includegraphics[width=12.5cm]{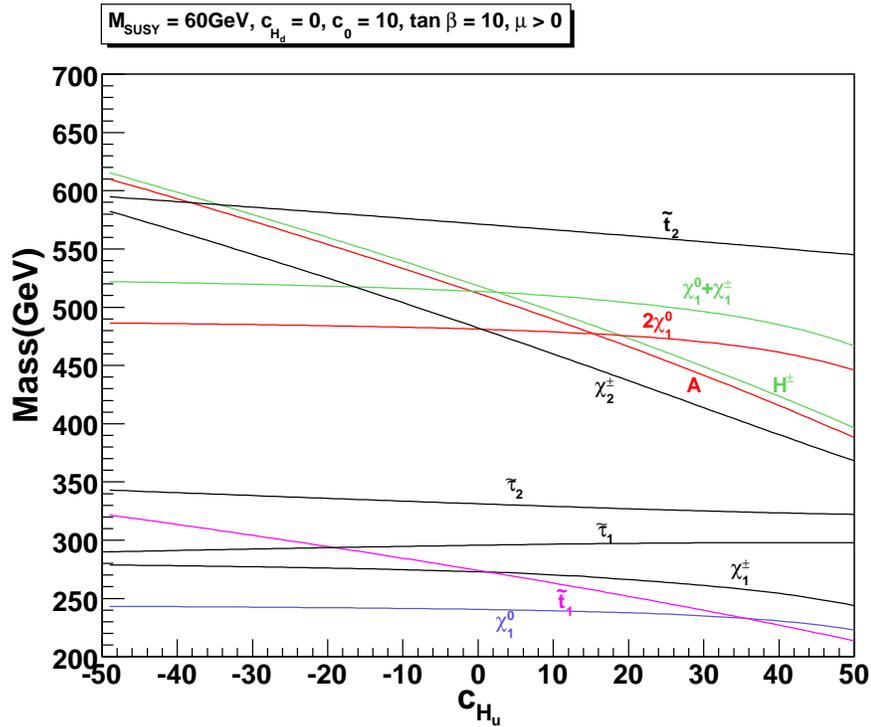} 
  \end{center}
\vspace{-0.5cm}
\caption{ \label{fig:0c} 
Left: the neutralino relic density in $M_{\rm SUSY}$ -- $c_{H_u}$ plane of the gauge messenger model 
with $c_0=10$, $c_{H_d}=0$, $\mu > 0$ and $\tan \beta = 10$. 
Blue dots represent the region in which the neutralino relic density is within WMAP range. 
Shaded regions are excluded by various constraints. 
Right: the dependence of various superpartner masses on $c_{H_u}$ for the choice of parameters corresponding to the plot on the left with 
$M_{\rm SUSY} = 60$ GeV.
}
\end{figure}

\begin{figure}[t]
  \begin{center}
  \hspace{-.5cm}
    \includegraphics[width=12.5cm]{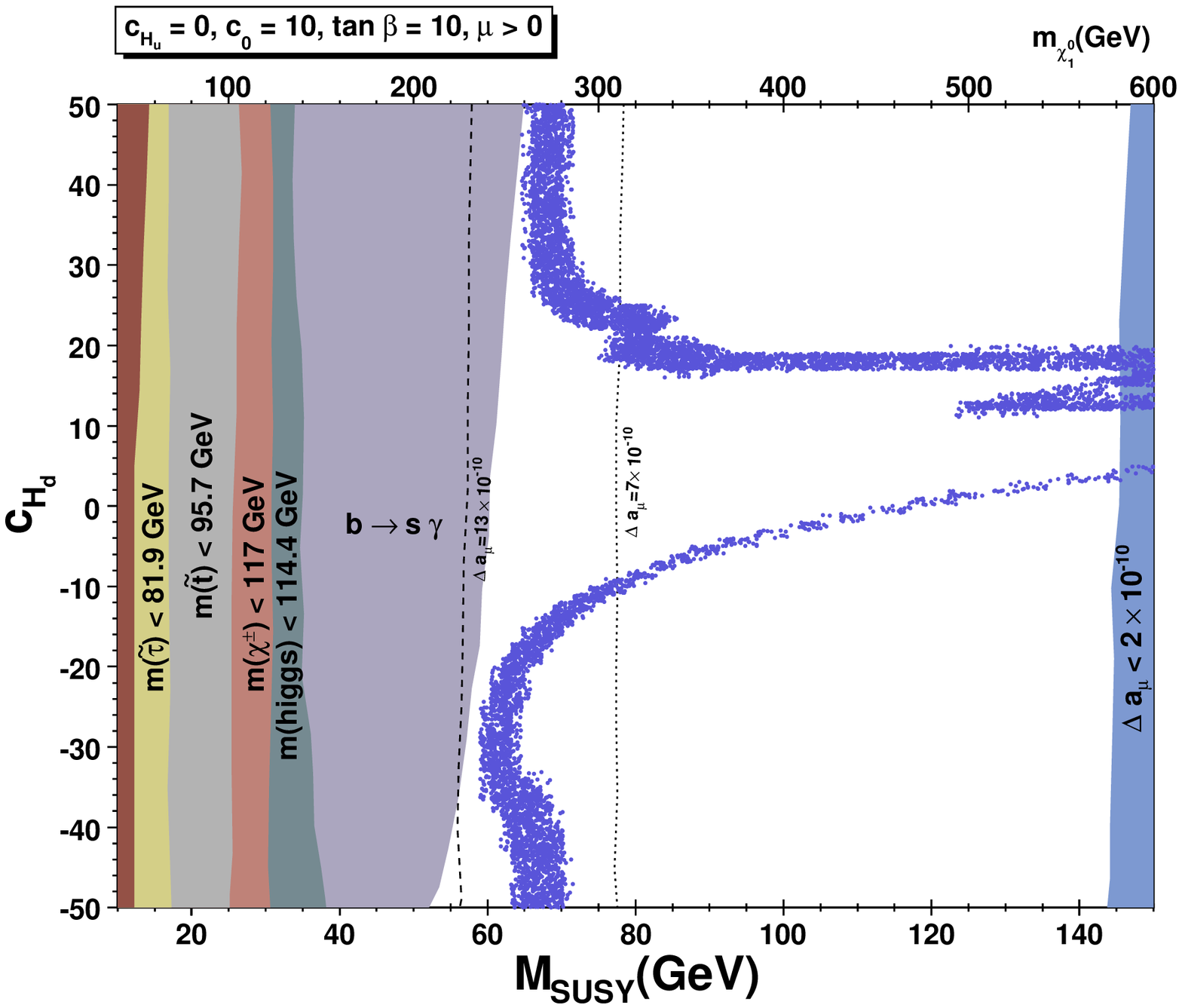} 
    \hspace{-.5cm}
  \includegraphics[width=12.5cm]{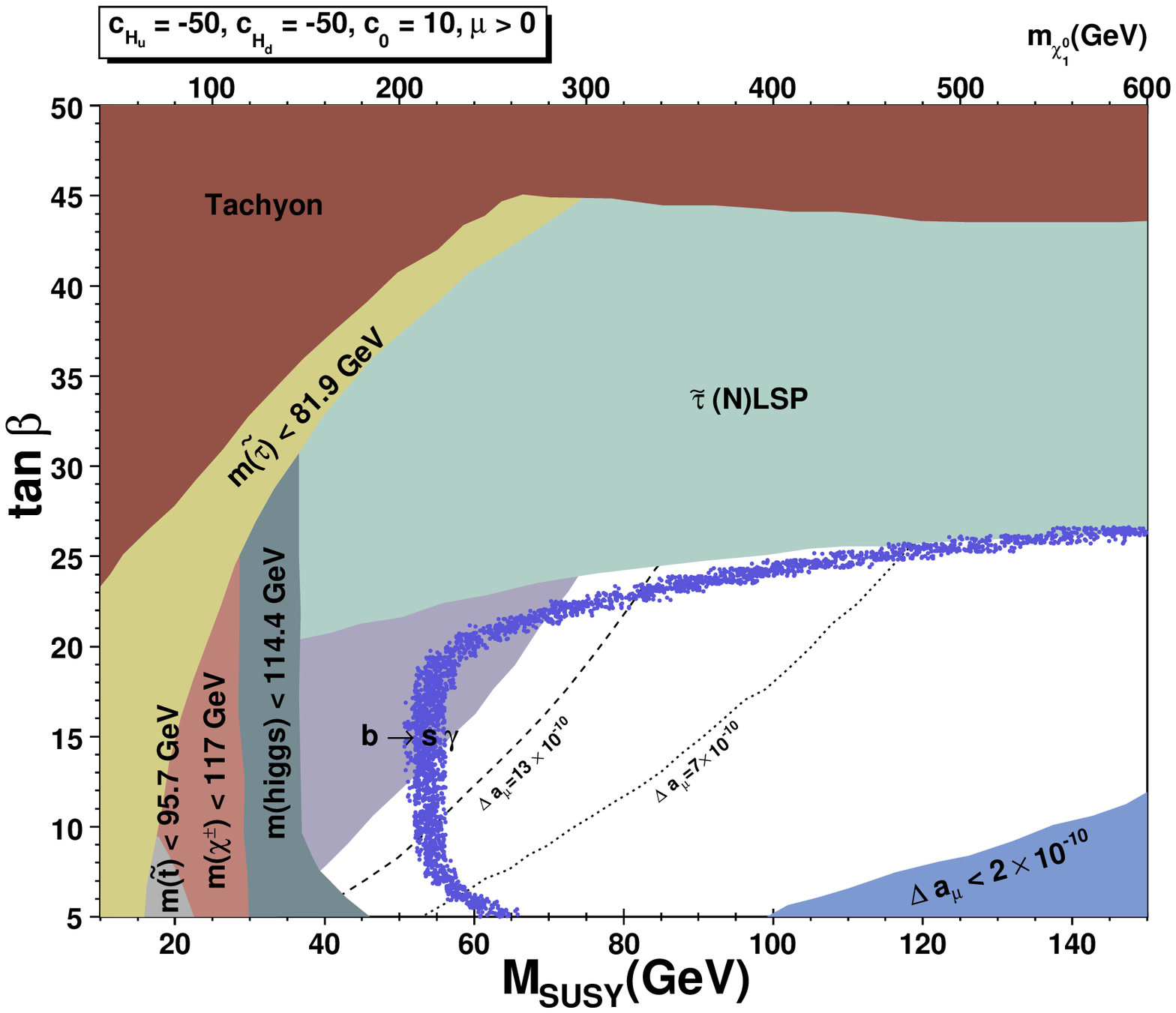}
  \end{center}
\vspace{-0.5cm}
\caption{ \label{fig:0d_3a} 
Left: the neutralino relic density in $M_{\rm SUSY}$ -- $c_{H_d}$ plane of the gauge messenger model 
with $c_0=10$, $c_{H_u}=0$, $\mu > 0$ and $\tan \beta = 10$. 
Right: the neutralino relic density in $M_{\rm SUSY}$ -- $\tan \beta$ plane of the gauge messenger model 
with $c_{H_u}=-50$, $c_{H_d}=-50$, $\mu > 0$ and $\tan \beta = 10$.
Blue dots represent the region in which the neutralino relic density is within WMAP range. 
Shaded regions are excluded by various constraints. 
}
\end{figure}

Results for the neutralino relic density in gauge messenger models are given in Figs.~\ref{fig:0a_0a'}~--~\ref{fig:0d_3a}. 
We start the discussion with 
 Fig.~\ref{fig:0a_0a'} in which we present the results for simple gauge messenger model with additional contribution to squark and slepton masses, $c_0 = 10$ (up) and $c_0=20$ (down). Additional contribution, e.g. from gravity, at this level is enough to push masses of all squarks and sleptons above the neutralino mass in a large region of the parameter space. Increasing $c_0$ shrinks the region of stau (N)LSP and opens up the region with neutralino (N)LSP.
Blue dots represent the region in which the neutralino relic density is within WMAP range.
 The top part of the blue band corresponds to the region of stau co-annihilation. This is easy to understand because the blue band stretches along the line dividing the neutralino and stau (N)LSP regions.
The bottom part of the blue band is due to the CP odd Higgs resonance which is not obvious from the plots but it will become clearer in later discussion. In most of the region for $c_0 = 10$ the co-annihilation with stop is also important and it is the dominant process in the region where the two bands meet. However, this does not mean that stop co-annihilation and thus the special choice of $c_0$ we made is crucial for obtaining the correct amount of the neutralino relic density. For $c_0=20$ the contribution from stop co-annihilation is no longer significant but the shape of the blue band is very similar only shifted to the left, to the region of smaller neutralino mass, in which the bino-wino-higgsino mixing and the chargino co-annihilation become important.

For even larger values of $c_0$, see   Fig.~\ref{fig:0b},   the effects coming from the exchange of or co-annihilation with squarks and sleptons disappear as squarks and sleptons become heavy and the band of the correct relic density is independent of $c_0$.
%~\footnote{Note that for large values of $c_0$ the boundary conditions for scalar masses resemble more and more mSUGRA boundary conditions. The generated mixing in the stop sector decreases with increasing $c_0$ and eventually it is no longer possible to satisfy the limit on the Higgs mass. Very large value of $c_0$ which leads to $\sim 1$ TeV stop masses would again satisfy the limit on the Higgs mass for any $M_{\rm SUSY}$.} 
The residual small $c_0$ dependence comes from the fact that increasing $c_0$ influences the renormalization group evolution of $m_{H_u}^2$ in such a way that the size of the $\mu$ term  increases which 
consequently reduces the mixture of higgsino and wino in the lightest neutralino. As a result, the correct value of the neutralino relic density is obtained with slightly lighter neutralino. 

Let us discuss the neutralino annihilation process in detail for one specific point
from Fig.~\ref{fig:0b} with 
$M_{\rm SUSY}=42$ GeV and $c_0=60$.
This point is away from the CP odd Higgs resonance and the relic density, $\Omega_{\rm DM} h^2 = 0.11$, reflects the composition of the lightest neutralino. 
The lightest neutralino is mostly bino with small mixtures of wino and higgsinos:
\bea
N_{11} & = & 0.95, \quad N_{12} = -0.22, \quad N_{13} = 0.18, \quad N_{14} = -0.09. \nn
\eea
The lightest and the next-to-lightest neuralinos and the light chargino are nearly degenerate:
\bea
m_{\chi^0_1} & = & 167 \ {\rm GeV}, \nn \\
m_{\chi^0_2} & = & 193 \ {\rm GeV}, \nn \\
m_{\chi^+_1} & = & 191 \ {\rm GeV}.
\eea
The dominant annihilation channel for this point is $\chi^0_1 \chi^0_1 \rightarrow W^+ W^-$ which represents 31\% of the annihilation cross section at the freezeout temperature.
It is mediated by the t-channel exchange of charginos and thus the wino component of $\chi^0_1$ plays an important role since the light chargino is mostly the wino.
Also  important channel is $\chi^0_1 \chi^0_1 \rightarrow b \bar{b}$ which contributes 24\%
indicating that the CP odd Higgs mediated s-channel diagram makes a contribution even away from the resonance. This is again a consequence of the wino and higgsino mixing (the higgsino-bino-A and higgsino-wino-A interactions are crucial). The amplitude for this process scales as $N_{14} (N_{12}- \tan \theta_W N_{11})$ and with $N_{12} \sim -0.2$ we see that the wino component enhances this process by $\sim 60 \%$.
Finally, slepton mediated t-channel diagrams contribute less than 10\%.

Chargino co-annihilation is always present since in the gauge messenger model the wino (and thus the lightest chargino) is only about 10\% heavier than the bino.
The chargino co-annihilation for this point contributes about 20\% to the annihilation cross section at the freezeout temperature. 
It is mediated mainly by the W boson in the s-channel which contributes about 10\% and also, to a smaller extent, by the charged Higgs in the s-channel.

In summary, the wino and higgsino mixing and the chargino co-annihilation play an important role 
in obtaining the correct amount of the neutralino relic density in gauge messenger models in the region with fairly light superpartners (not ruled out by direct searches or the limit on the Higgs boson mass). With this knowledge we can continue with the discussion of more typical (and more complex)  scenarios when additional co-annihilations and/or resonances further enhance the neutralino annihilation cross section.

In Fig.~\ref{fig:0c} (up)  we study the dependence of the neutralino relic density on the additional contribution to the mass squared for $H_u$.
To better understand the behavior of the neutralino relic density we also plot the dependence of SUSY spectrum on $c_{H_u}$ for fixed value of $M_{\rm SUSY}$ in Fig.~\ref{fig:0c} (down). For $c_{H_u} \gtrsim 25$ the lightest stop mass is very close to the lightest neutralino mass and the stop co-annihilation is dominant. The correct amount of the relic density is then obtained in an almost horizontal band at $c_{H_u} \simeq 30$. Going to smaller $c_{H_u}$ the difference between stop and neutralino masses is increasing and the co-annihilation with stop is no longer important. The CP odd Higgs resonance takes over for  $c_{H_u} \simeq 20$ at $M_{\rm SUSY} = 60$ GeV and somewhat smaller $c_{H_u}$ for larger $M_{\rm SUSY}$. The second smaller peak is mainly due to co-anihilation with the lightest chargino through  the charged Higgs resonance which happens when $m_{H^+} \simeq  m_{\chi^0_1} + m_{\chi^+_1}$ and to a smaller extent due to co-anihilation with the  second lightest neutralino through  the CP odd Higgs resonance which happens when   $m_{A} \simeq  m_{\chi^0_1} + m_{\chi^0_2}$. Since the lightest chargino and the second lightest neutralino are mostly winos these two resonances happen in the same region,  $c_{H_u} \simeq 0$ at $M_{\rm SUSY} = 60$, and  continue to somewhat smaller $c_{H_u}$ for larger $M_{\rm SUSY}$. 
Finally, decreasing  $c_{H_u}$ further takes the lightest neutralino away from stop co-annihilation and resonance regions and the blue band of the correct relic density is almost vertical in this region. The residual $c_{H_u}$ dependence comes from the fact that $c_{H_u}$ changes the size of the $\mu$ term which then varies the mixture of higgsino and wino in the lightest neutralino. The correct amount of the neutralino relic density in this region is obtained entirely due to the wino and higgsino mixing and the chargino co-annihilation as discussed in the example above.

The dependence of the neutralino relic density on the additional contribution to the mass squared for $H_d$
is given in Fig.~\ref{fig:0d_3a} (up). The $c_{H_d}$ controls masses of the heavy CP even, charged and CP odd  Higgs bosons and only negligibly affect everything else. Thus the region of the correct relic density is a vertical band except for the CP odd and charged Higgs resonances. Finally, in Fig.~\ref{fig:0d_3a} (down) we chose such values of $c_{H_u}$ and $c_{H_d}$ that the CP odd Higgs resonance does not appear. This plot is similar to those in Fig.~\ref{fig:0a_0a'}, but now the stau co-annihilation region turns into a vertical band signaling independence on $\tan \beta$. Similar vertical band appears also in the mSUGRA scenario, see Fig.~\ref{fig:mSUGRA_m0=m12}, but it is in the region ruled out by direct searches for SUSY and the Higgs boson.

\subsection{Discussion of $b \to s \gamma$ and  muon $g-2$}

From Figs.~\ref{fig:0a_0a'}~--~\ref{fig:0d_3a} we see that the limits on 
$B(b \rightarrow s \gamma)$ are typically the most constraining out of all 
direct and indirect limits.
The charged Higgs contribution is additive to the standard model contribution and scales as
\bea                                                                                 
B(b \rightarrow s \gamma)_{H^\pm} & \propto & \f{m_t^2}{m_{H^\pm}},                             
\eea
while the chargino-stop loop contributes as
\bea                                                                                 
B(b \rightarrow s \gamma)_{\tilde{t}} & \propto & \f{\mu A_t \tan                    
\beta}{m_{\tilde{t}}^2}.                                                   
\eea 
The chargino-stop loop contributes with opposite sign compared to the charged Higgs diagram if 
$\mu A_t$ is negative.
In gauge messenger models the charged Higgs is typically heavier than stop 
and the chargino-stop loop dominates the new physics contribution.
As a result,
the predicted branching ratio becomes lower than the Standard Model result and             
the lower bound on $B(b \to s \gamma)$ plays an important role.

In the limit when $M_1$, $M_2$, $m_{\tilde {\mu}}$ and          
$ m_{{\tilde \nu}_{\mu}}$ are approximately equal, which is the case in gauge messenger models,
and $\mu > M_1, \ M_2$, the expression of the supersymmetric contribution to muon
anomalous magnetic moment~\cite{Moroi:1995yh} simplifies to:
\bea                                                                                 
\Delta a_{\mu}^{\rm SUSY} & \simeq & \f{g_2^2 m_{\mu}^2}{32\pi^2} \f{\mu             
M}{M^2(\mu^2-M^2)} \tan \beta,                                                       
\eea   
where $M$ represents sneutrino, smuon, chargino or neutralino masses.
It can be rewritten as                                                            
\bea                                                                                 
\Delta a_{\mu}^{\rm SUSY} & \simeq & 13 \left( \f{100 {\rm GeV}}{M}                  
\right)^2 \left( \f{ \mu M}{\mu^2 - M^2} \right)\tan \beta \times                    
10^{-10}.                                                                            
\eea                                                                                                                                                                      
As a result, we obtain a relation between $M$ 
and $\tan \beta$. In most of the parameter              
space $\mu$ is just about twice as large as the lightest neutralino mass and thus we can             
set $ \f{ \mu M}{\mu^2 - M^2} \simeq \f{2}{3}$ in which case we get
\bea                                                                                 
\Delta a_{\mu} \times 10^{10} & \simeq & \f{26}{3} \left( \f{100 {\rm                
GeV}}{M} \right)^2 \tan \beta.                                                       
\eea                                                                                 
 
Assuming conservative bounds $2 \times 10^{-10} < \Delta a_{\mu} < 50 \times 10^{-10}$ a discussed in Sec.~\ref{sec:constraints} we can derive the lower and upper bounds on $M$ as a function of $\tan \beta$:
\bea                                                                                 
M_{\rm lower} & \sim & 40 \sqrt{ \tan \beta} \ {\rm GeV},                          
\eea                                                                                 
and                                                                                  
\bea                                                                                 
M_{\rm upper} & \sim & 200 \sqrt{ \tan \beta} \ {\rm GeV}. 
\eea                                                                                                                                                                      
For $\tan \beta =10$ we find $130 \ {\rm GeV} \lesssim  M \lesssim 630$ GeV and 
similarly for $\tan \beta = 50$ we have $280 \ {\rm GeV} \lesssim  M \lesssim 1400$ GeV.
In Figs.~\ref{fig:0a_0a'}~--~\ref{fig:0d_3a} the value of $M$ approximately corresponds to the neutralino mass represented by the top axis.
It is interesting to note that the indirect bound from the upper limit on the            
muon anomalous magnetic moment is already well above the direct search               
limits on superpartners. 

As a result of the squeezed spectrum of gauge messenger models the limits on 
$B(b \rightarrow s \gamma)$ and the muon $g-2$ are almost parallel to each other, see Figs.~\ref{fig:0a_0a'}~--~\ref{fig:0d_3a}. This is a consequence of the SUSY contribution to both processes scaling approximately as $\tan \beta/ M^2$.
The limits on $B(b \rightarrow s \gamma)$ constrain the SUSY spectrum from below while the limits on  $g-2$ constrain the parameter space from above. The allowed parameter space is then only a strip in between these two bounds. This is a characteristic feature of models with squeezed spectrum.
If the required value of $\Delta a_{\mu}$ turns out to be close to the upper range of current estimates
most of the parameter space of gauge messenger models we considered will be ruled out with only tiny regions remaining, see Figs.~\ref{fig:0a_0a'}~--~\ref{fig:0d_3a}. Interestingly, it is still possible to obtain the correct amount of the dark matter density in these tiny regions, see Figs.~\ref{fig:0c} and \ref{fig:0d_3a} (down).

\subsection{Direct Detection of Neutralino Dark Matter}

In this section we calculate the 
spin dependent and spin independent nuetralino-nucleon cross sections in gauge messenger models.
\begin{figure}[t]
  \begin{center}
\subfigure[Spin dependent neutralino-nucleon cross section
for simple gauge messenger with $c_0=10$]{
\includegraphics[width=9.5cm]{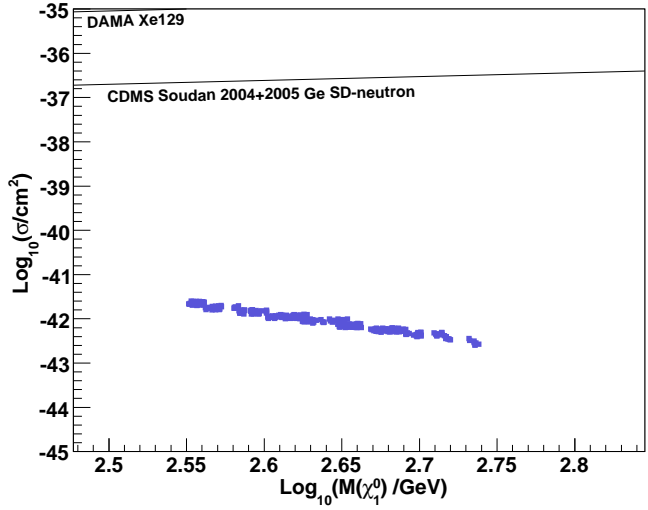}}
\subfigure[Spin independent neutralino-nucleon cross section
for simple gauge messenger with $c_0=10$]{
  \includegraphics[width=9.5cm]{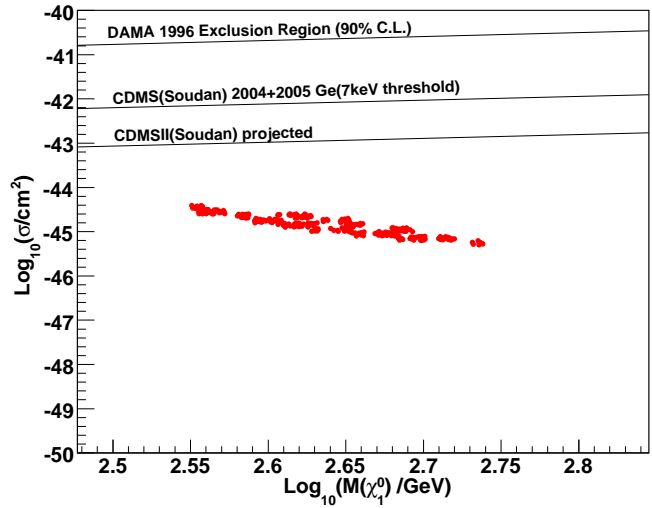}}
  \end{center}
\caption{ \label{fig:detection_c0=10} 
Spin dependent and spin independent neutralino-nucleon cross sections for points with the correct neutralino relic density from Fig.~\ref{fig:0a_0a'} (left), the gauge messenger model 
with $c_{H_u}=c_{H_d}=0$, $\mu > 0$ and
$c_0 = 10$, that satisfy all direct and indirect constraints. 
The lines represent
the current CDMS limits~\cite{Akerib:2005kh} and expected limits from CDMSII~\cite{Brink:2005ej} for spin independent cross section. 
}
\end{figure}
\begin{figure}[t]
  \begin{center}
\subfigure[Spin dependent neutralino-nucleon cross section
for the extended parameter scan]{
  \includegraphics[width=9.5cm]{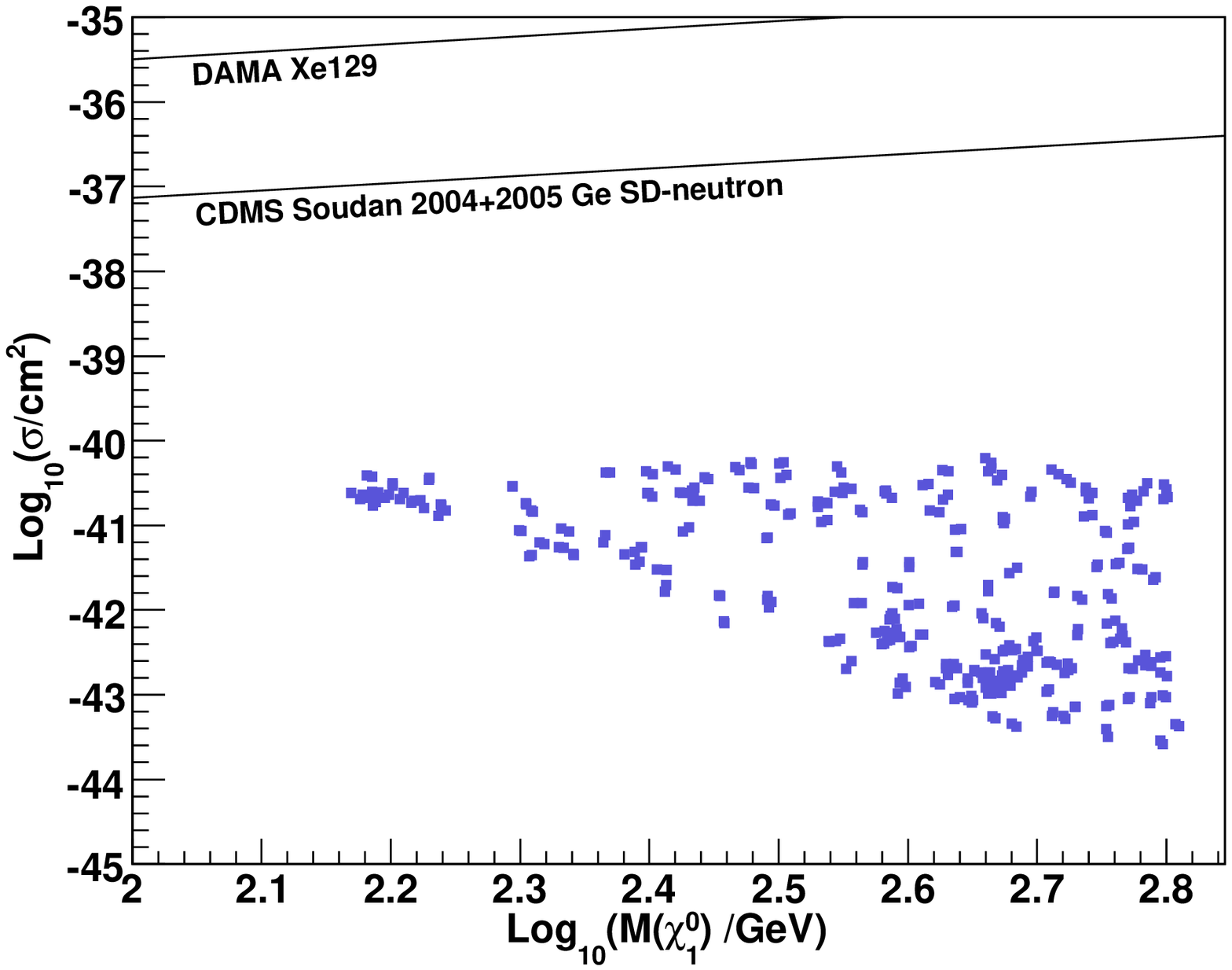}}
\subfigure[Spin independent neutralino-nucleon cross section
for the extended parameter scan]{
  \includegraphics[width=9.5cm]{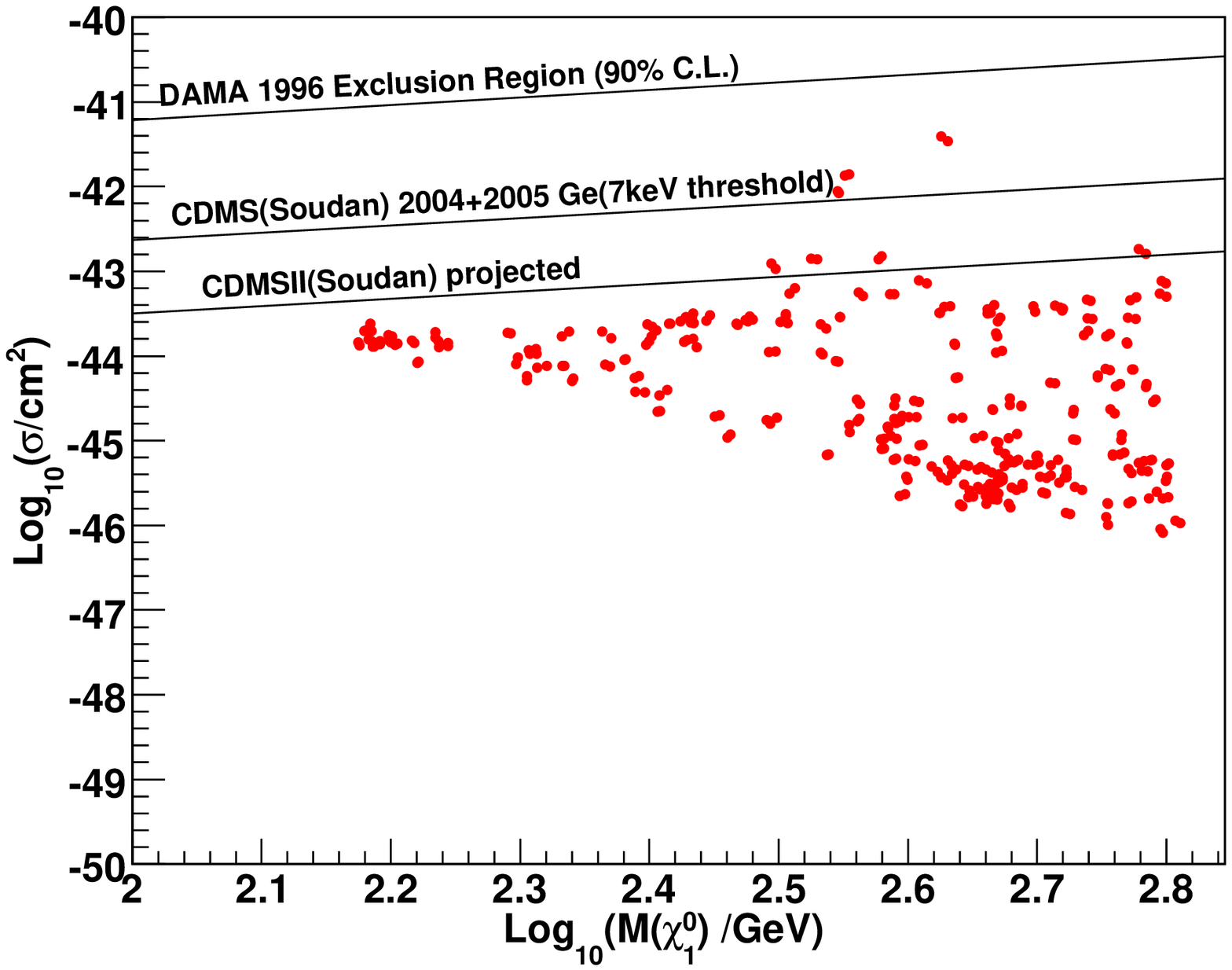}}
  \end{center}
\caption{ \label{fig:detection_all} 
Spin dependent and spin independent neutralino-nucleon cross sections for points with the correct neutralino relic density satisfying all direct and indirect constraints obtained in an extended scan over whole parameter space of gauge messenger models discussed in this paper. The lines represent
the current CDMS limits and expected limits from CDMSII for spin independent cross section.
 }
\end{figure}
Spin independent neutralino-nucleon cross section is dominated by (light and heavy) Higgs mediated t-channel diagrams which are controled by the higgsino component of the lightest neutralino:
\bea
\sigma_{\chi N} & = & \f{g^2 {g^\prime}^2 m_N^4}{4\pi M_W^2} \left[ - \f{X_d \tan \beta N_{13}}{m_H^2} + \f{X_u N_{14}}{m_h^2} \right]^2,
\eea 
where $X_d = f_{T_s} + \f{2}{27} f_{TG}$ and $X_u = f_{T_u} + \f{4}{27} f_{TG}$~\cite{Bertone:2004pz}. 

Substituting $N_{13}$ and $N_{14}$ from Eqs.~(\ref{eq:N13}) and (\ref{eq:N14}) we find
\bea
\sigma_{\chi N} & = & \f{{g^\prime}^4 m_N^4}{4\pi} \left(\f{\mu^2}{(\mu^2-M_1^2)^2}\right)  \left[ \f{X_d \tan \beta}{m_H^2} + (\f{M_1}{\mu} )\f{(X_u+X_d)}{m_h^2} \right]^2.
\eea 
Squark exchange diagrams are negligible due to the hypercharge as long as the squark masses are comparable to the heavy Higgs mass.
Inserting the numbers $X_u = 0.144$ and $X_d = 0.18$~\cite{Bertone:2004pz}, we get the direct detection rate close to the one we obtained using DarkSUSY.

The detection cross sections  for points with the correct neutralino relic density from Fig.~\ref{fig:0a_0a'} (up), the gauge messenger model 
with $c_{H_u}=c_{H_d}=0$, $\mu > 0$ and
$c_0 = 10$, that satisfy all direct and indirect constraints are given in Fig.~\ref{fig:detection_c0=10}.
In gauge messenger models with no additional contribution to Higgs soft masses and only small contribution to other scalar masses enough to make them heavier than the lightest neutralino the direct detection cross section scales as $\sigma_{\chi N} \propto \tan^2 \beta / M_{\rm SUSY}^6$ for $\tan \beta \ge 10$. This behavior is clearly visible in Fig.~\ref{fig:detection_c0=10}. The thickness of the line is determined by the allowed region for $\tan \beta$, in this case $5 < \tan \beta <  25$, see Fig.~\ref{fig:0a_0a'}.
 The predicted cross sections are not within the reach of CDMSII~\cite{Brink:2005ej}.
Assuming additional contributions to  Higgs soft masses allows for a wider range of the higgsino and wino mixing and the range of the predicted detection cross sections spreads as is shown~Fig.~\ref{fig:detection_all}. Part of the parameter space is within the reach of CDMSII and the whole parameter space of gauge messenger models can be explored at Super-CDMS~\cite{Brink:2005ej}.

\section{Conclusions \label{sec:conclusions}}

The lightest neutralino in gauge messenger models is mostly the  bino with a sizable mixture of the wino and higgsino. 
The wino and higgsino components enhance the neutralino annihilation cross section. Furthermore,
the splitting between the bino and wino masses is at the level of 10\% and thus the co-annihilation with the chargino is contributing in the whole region of the parameter space. These two features, the lightest neutralino being a mixture of  the bino, wino and higgsino, and the chargino co-annihilation, are sufficient for obtaining the correct neutralino relic density to explain WMAP results with fairly light neutralino (and other superpartners) while satisfying all the constraints from    direct searches for superpartners and the limit on the Higgs boson mass.

This is in contrast with scenarios with the usual hierarchical spectrum, e.g. mSUGRA, in which the properties of the lightest neutralino (being bino-like) typically lead to the correct neutralino relic density in the region which is already ruled out by  direct SUSY and Higgs searches or disfavored by $b \to s \gamma$. 
In mSUGRA-like models  obtaining the correct amount of the neutralino relic density relies on special co-annihilation and resonance regions which are critically sensitive to small variations of independent parameters. Due to a large hierarchy in the spectrum these surviving strips are typically well separated by large regions of the parameter space ruled out by WMAP data. 
 
In  gauge messenger models, as a result of the squeezed spectrum of superpartners,
 various co-annihilation and resonance regions overlap
and very often the correct amount of the neutralino relic density is generated as an interplay of several processes. For example the stop co-annihilation contributes significantly  in a large region of the parameter space. This can be easily understood from the fact that both stop and neutralino masses are mainly controled by the same parameter and as it happens the neutralino and stop masses are very close to each other. Varying contributions to scalar masses from other sources is only slowly changing this relation. Furthermore,
even if we increase stop masses by assuming an independent additional contribution, which effectively shuts down the stop co-annihilation,  the band of the correct neutralino relic density only moves
to the region with somewhat lighter neutralino which still satisfies the limits from direct SUSY and Higgs searches.   
This feature makes the explanation of the observed amount of the dark matter density much less sensitive to fundamental parameters. 

In gauge messenger models with no additional contribution to Higgs soft masses and only small contribution to other scalar masses enough to make them heavier than the lightest neutralino the direct detection cross section is predicted to be in the range $10^{-46} \ {\rm cm}^2$ -- $10^{-44} \ {\rm cm}^2$ which is   not within the reach of CDMSII but can be explored at Super-CDMS.

Some of the results concerning the neutralino relic density in gauge messenger models, namely the presence of various co-annihilation regions, originate from the sqeezed SUSY spectrum. Therefore we expect similar results for  other models derived in  different contexts which lead to  
squeezed spectrum, e.g. deflected anomaly mediation \cite{Pomarol:1999ie}
\cite{Rattazzi:1999qg} \cite{Cesarini:2006jp} 
and mirage mediation 
\cite{Choi:2005uz} \cite{Choi:2005hd} \cite{Kitano:2005wc}~\cite{Choi:2006im}.
However, the special features of the gauge messenger model related to the bino-wino-higgsino mixed dark matter and with that associated chargino co-annihilation depend on details of a model and is not automatically guaranteed by the squeezeness. 

In conclusion, let us note that both natural EWSB and natural explanation of the correct amount of the dark matter density independently disfavor models with hierarchical spectrum. 
Models with squeezed spectrum seem to be favored and thus it is desirable to explore their phenomenological and collider predictions.

\section*{Acknowledgement}

RD thanks the Aspen Center for Physics for hospitality and support during the course of
this research.
HK thanks Harvard Univeversity theory group for hospitality during his
visit and IW thanks the Ohio State University group during his visit. 
RD is supported in part by U.S. Department of Energy, grant
number DE-FG02-90ER40542. HK and IW are supported by the ABRL Grant No.
R14-2003-012-01001-0,
and KB and HK by the BK21 program of Ministry of Education,
and HK by Korea and the Science Research Center Program of the Korea Science
and Engineering Foundation through the Center for Quantum Spacetime
(CQUeST) of Sogang University with grant number R11-2005-021.

\section*{Appendix}

In this appendix we set conventions and derive  approximate formulae for the composition 
of the lightest neutralino in the gauge messenger model which are useful for
the discussion  of neutralino relic density.

The neutralino mass matrix in the basis (B, W, $h_d$, $h_u$) is given by:
\begin{eqnarray}
M_N  =   \left(
\begin{array}{cccc}
M_1 & 0 & -M_Z c_\beta s_{\theta_W} & M_Z s_\beta s_{\theta_W} \\
0 & M_2 & M_Z c_\beta c_{\theta_W} & -M_Z s_\beta c_{\theta_W} \\
-M_Z c_\beta s_{\theta_W} & M_Z c_\beta c_{\theta_W} & 0 & -\mu \\
M_Z s_\beta s_{\theta_W} & -M_Z s_\beta c_{\theta_W} & -\mu & 0
\end{array}
\right), 
\end{eqnarray}
where $s_{\theta_W} \equiv \sin \theta_W$,  $c_{\theta_W} = \cos \theta_W$ with $\theta_W$ being the Weinberg angle (weak mixing angle) and similarly $s_\beta = \sin \beta$, $c_\beta = \cos \beta$ where $\tan \beta = \f{v_u}{v_d}$.

In the gauge messenger model, bino and wino masses are comparable. Thus it is convenient to express the wino mass in terms of the bino mass and a small parameter describing the difference,
\begin{eqnarray}
M_2 & = & M_1 (1 + \epsilon) .
\end{eqnarray}
Numerically $\epsilon \simeq 0.09$ and it is  almost independent of $\tan \beta$. Thus, for $M_1 < |\mu|$  the lightest neutralino is mostly bino
and the splitting between the bino and the wino is  at the level of $10~\%$. 

The neutralino mass matrix can be brought to a diagonal form by an orthogonal transformation,
\bea
M_{\rm diag} & = & N M_N N^T,
\eea
where $N_{1j}$, $j = 1,2,3$ and $4$ represent the mixture of B, W, $h_d$, $h_u$ in the lightest neutralino mass eigenstate.

In order to calculate $N_{1j}$, it is convenient to rotate the neutralino mass matrix to a basis in which the lower right 
$2 \times 2$ block is diagonal,
{\small
\begin{eqnarray}
\hspace{-2.5cm} M & =  & \left(
\begin{array}{cccc}
M_1 & 0 & -\f{1}{\sqrt{2}} M_Z s_{\theta_W} (s_\beta+ c_\beta) & \f{1}{\sqrt{2}}M_Z s_{\theta_W} (s_\beta - c_\beta) \\
0 & M_2 & \f{1}{\sqrt{2}} M_Z c_{\theta_W} (s_\beta+ c_\beta) & -\f{1}{\sqrt{2}}M_Z c_{\theta_W} (s_\beta - c_\beta) \\
-\f{1}{\sqrt{2}}M_Z  s_{\theta_W}  (s_\beta+ c_\beta) & \f{1}{\sqrt{2}}M_Z  c_{\theta_W}  (s_\beta+ c_\beta)& \mu & 0 \\
\f{1}{\sqrt{2}}M_Z s_{\theta_W}  (s_\beta- c_\beta) & -\f{1}{\sqrt{2}}M_Z c_{\theta_W}  (s_\beta- c_\beta)& 0 & -\mu
\end{array}
\right), \nonumber
\end{eqnarray}
}
which is obtained by an orthogonal transformation,
\bea
M & = & U M_N U^T, 
\eea
with
\bea
U & = & \left(
\begin{array}{cccc}
1 & 0 &0 & 0 \\
0 & 1 & 0 & 0 \\
0 & 0 & \f{1}{\sqrt{2}} &  -\f{1}{\sqrt{2}} \\
0 & 0 &  \f{1}{\sqrt{2}} &  \f{1}{\sqrt{2}}
\end{array}
\right). 
\label{eq:U}
\end{eqnarray}
The matrix $M$ can be diagonalized by an orthogonal transformation, 
\bea
M_{\rm diag} = & = & V M V^T.
\eea
The advantage of  $M$ is that we can treat off-diagonal elements as perturbations and calculate eigenvectors
(elements of $V$) using matrix perturbation formalism. 
Then, the mixing matrix $N$ (the diagonalization matrix in the original basis) is simply given as
\bea
N = V U.
\label{eq:N}
\eea

In the leading order, neglecting the off-diagonal elements of $M$,
 the diagonalization matrix $V$ is an identity matrix.
When the mass differences between eigenvalues are not extremely small, $(M_2 - M_1) \mu^2  \ge M_1 M_Z^2$,
or equivalently $\epsilon \mu^2 \ge M_Z^2$, non-degenerate perturbation formalism can be applied.
At the first order of perturbation theory we have:
\bea
V^{(1)}_{nm} & = & \f{M_{mn}}{\Delta_{nm}},
\eea
where $\Delta_{mn} = M_{mm} - M_{nn}$. Similarly, the second order corrections are given as:
\bea
V^{(2)}_{nl} & = & \sum_{m \neq n} \f{M_{lm} M_{mn}}{\Delta_{nl} \Delta_{nm}}.
\eea
For  $M_1 < |\mu|$ we find:
\bea
V^{(1)}_{11} & = & 0, \\
V^{(1)}_{12} & = & 0, \\
V^{(1)}_{13} & = & - \f{M_{31}}{\mu - M_1} = \f{M_Z \sin \theta_W (\sin \beta + \cos \beta)}{\sqrt{2}(\mu - M_1)}, \\
V^{(1)}_{14} & = & + \f{M_{41}}{\mu + M_1} = \f{M_Z \sin \theta_W ( \sin \beta - \cos \beta)}{\sqrt{2}(M_1+\mu)},
\eea
and thus the higgsino component  in the lightest neutralino appears at the first order.

Since $V^{(1)}_{12} = 0$ it is necessary to calculate the contribution from the next order.
This contribution 
is small in general but can significantly alter the result when $M_1 \sim M_2$.
The second order correction is
\bea
V^{(2)}_{12} & = & -\f{M_{23} M_{31}}{(M_2 - M_1)(M_1 - \mu)}
-\f{M_{24} M_{41}}{(M_2 - M_1)(M_1 + \mu)}, \nn \\
& = & -\f{M_Z^2 \sin 2\theta_W (\sin \beta +\cos \beta)^2}{4 \epsilon M_1 (\mu - M_1)}
 +\f{M_Z^2 \sin 2\theta_W (\sin \beta -\cos \beta)^2}{4 \epsilon M_1 (\mu + M_1)},
\eea
and,
since $\epsilon \sim 0.1$, it is comparable to the first order corrections coming from the higgsino mass.
Therefore, we have a sizable bino-wino mixing in addition to bino-higgsino mixing.

The diagonalization matrix $V$ is then approximately given as $V \simeq 1 + V^{(1)} + V^{(2)}$. 
Finally, we can find the components of the mixing matrix in the original interaction basis. Using Eqs.~(\ref{eq:N}) and (\ref{eq:U}) we get:
\bea
N_{11} & \simeq & 1, \\
N_{12} & \simeq & V^{(2)}_{12}, \\
N_{13} & = & +\f{1}{\sqrt{2}} V_{13} + \f{1}{\sqrt{2}} V_{14} \simeq  \f{M_Z \sin \theta_W (\mu \sin \beta  + M_1 \cos \beta )}{\mu^2 - M_1^2},\\
N_{14} & = &-\f{1}{\sqrt{2}} V_{13} + \f{1}{\sqrt{2}} V_{14} \simeq -\f{M_Z \sin \theta_W (M_1 \sin \beta + \mu \cos \beta)}{\mu^2 - M_1^2}.
\eea
For  $\tan \beta \ge 10$ these formulae can be further simplified:
\bea
N_{11} & \simeq & 1, \\
N_{12} & \simeq & -\f{M_Z^2 \sin 2\theta_W}{2\epsilon (\mu^2 - M_1^2)}, \\
N_{13} & \simeq & +\f{\mu M_Z \sin \theta_W }{\mu^2 - M_1^2}, \\
N_{14} & \simeq & -\f{M_1 M_Z \sin \theta_W }{\mu^2 - M_1^2}.
\eea

\end{document}